# A new hybrid spectral similarity measure for discrimination of *Vigna* species

M.NARESH KUMAR*†, M.V.R SESHASAI†, K.S VARA PRASAD‡,
V.KAMALA‡, KV RAMANA†, P.S. Roy†

†National Remote Sensing Centre, Balanagar Hyderabad

‡National Bureau of Plant genetic Resources (NBPGR), Hyderabad

**Abstract**
The reflectance spectrum of the species in a hyperspectral data can be modelled as an *n*-dimensional vector. The spectral angle mapper computes the angle between the vectors which is used to discriminate the species. The spectral information divergence models the data as a probability distribution so that the spectral variability between the bands can be extracted using the stochastic measures. The hybrid approach of spectral angle mapper and spectral information divergence is found to be better discriminator than spectral angle mapper or spectral information divergence alone. The spectral correlation angle is computed as a cosine of the angle of the Pearsonian correlation coefficient between the vectors. The spectral correlation angle is a better measure than the spectral angle mapper as it considers only standardized values of the vectors rather than the absolute values of the vector. In the present paper a new hybrid measure is proposed which is based on the spectral correlation angle and the spectral information divergence. The proposed method has been compared with the hybrid approach of spectral information divergence and spectral angle mapper for discrimination of crops belonging to *Vigna* species using measures like relative spectral discriminatory power, relative discriminatory probability and relative discriminatory entropy in different spectral regions. Experimental results using the laboratory spectra show that the proposed method gives higher relative discriminatory power in 400nm-700nm spectral region.
*Keywords*: Spectral information divergence, Spectral correlation angle, Spectral angle mapper, Relative spectral discriminatory probability, Relative spectral discriminatory entropy, Relative spectral discriminator power, and *Vigna* species.

## 1 Introduction

The discrimination of targets is based on the comparison of the given spectra with the reference spectra available as end-members in a spectral library. The comparison is done using the similarity as a criterion (Chang 2000, Du et al 2004, Farifteh et al 2006, Van der Meer 2006). The spectral angle mapper represses the influence of shading to enhance the target reflectance because of which it has been extensively used for discrimination of targets like plant species (Bakker et al 2002, Clark 2005, Clark et al 1990, 1999). Stochastic measures such as spectral information divergence consider the spectral band-to-band variability as a result of uncertainty incurred by randomness. The spectrum can be modelled as a probability distribution so that the spectral properties can be further described by statistical moments of any order (Chang 2000). The hybrid approaches of spectral angle mapper and spectral information divergence is found to increase the discriminatory power as against the individual measures (Du et al 2003, 2004).

---

* Corresponding author. Email: nareshkumar_m@nrsc.gov.in

The spectral angle mapper has a limitation that it cannot distinguish between negative and positive correlations as only the absolute value is considered. The spectral correlation angle on the other hand eliminates the negative correlation and maintains the spectral angle mapper characteristics of minimizing the shading effect resulting in better results. In this paper the hybrid measure of spectral correlation angle and spectral information divergence is proposed and is compared with the hybrid measure of spectral angle mapper and spectral information divergence to discriminate the crop species blackgram, greengram, horsegram, cowpea, and ricebean belonging to *Vigna* genus. Formulae for different spectral similarity and discriminatory are presented by different authors (Van der Meer 2006, Chang 2000), the same has presented in section 3 of this paper for the benefit of the readers.

The Objectives of this paper were
(1) to investigate and quantify the spectral reflectance of crop species belonging to *Vigna* genus;
(2) to create a methodology for discriminating crop species belonging to *Vigna* genus using hyperspectral data;
(3) to develop mathematical formulation for a new hybrid spectral similarity measure based on spectral correlation angle and spectral information divergence;
(4) to evaluate discriminatory powers of the hybrid measures spectral angle mapper, spectral information divergence and spectral correlation angle, spectral information divergence in different spectral regions; and
(5) to develop a decision table suggesting the reference spectra, spectral range and hybrid spectral measure to be used for discriminating the crop species of *Vigna* genus.

## 2 Data collection protocols and texture analysis

### 2.1 *Protocol for Spectral Collection*

The spectral reflectance of the selected crops belonging to the genus *Vigna* were measured using FieldSpec Pro Spectroradiometer FR (350nm-2500 nm) of Analytical Spectral devices (ASD), a hand-held, multi-band ground truth radiometer operating in three wavelength regions spread across 350nm to 2500nm. The spectroradiometer has three internal diodes to measure the radiation, fixed at 350nm-1050nm, 900nm-1850nm and 1700nm-2500nm. Integration time is set automatically for each of the three arrays to optimize the incoming radiation levels in all three regions.

The collection of spectral measurements include the optimization of the instrument for the integration time, measuring the dark current, collection of reflectance over white panel / spectralon panel followed by measurements of the target. Spectral reflectance measurements were made with 25º FOV sensor by keeping the instrument about one meter above the crop canopy with the sensor facing the crop and oriented normal to the plant. The observations were recorded on cloud free days at around solar noon time. Spectroradiometer was configured to average 25 samples per spectrum and spectral measurements of the selected crops were replicated 5 times. Reflectance observations over barium sulphate panel were collected at regular intervals of 15 minutes for referencing to account for variations in the solar illumination as a function of time. Due care was taken not to overcast shadow over the area being scanned. Windows based software VIEWSPEC PRO was used for post processing of the data collected.

## 2.2 Texture or Spectral responses

The crop species blackgram, greengram, horsegram, cowpea and ricebean all belonging to the same genus *Vigna*, have been considered for our study. The average reflectance of the crop species collected from the spectroradiometer in different spectral regions is shown in the (figure 1). The spectral regions full spectral region 400nm-2300nm (168 bands) consisting of visible region 400nm-700nm (31 bands), NIR region 700nm-1290nm (60 bands) and SWIR region 1510nm-2300nm (78 bands) at 10nm interval are considered for analysis. The reflectance measurements of the crop species is filtered in the spectral range 1800nm-2000nm and 1300nm-1500nm due to sensor noise.

[Include Figure 1 here]

Absorption in the red and green region is noticed due to the pigments. Red edge inflection around 700nm is observed, followed by a plateau region up to 1200nm is observed which is attributed to the internal cellular structure and the turgidity of the cells that influences the total internal reflections. Absorption due to water is noticed in around 1650nm and 2250nm. In 400nm-700nm spectral range ricebean has a distinct spectral profile whereas horsegram and greengram have similar spectral profiles. In 700nm-1290nm spectral range cowpea has distinct spectral profile whereas the ricebean and blackgram, horsegram and greengram have similar spectral profiles. In 1510nm-2300nm spectral range cowpea and blackgram have similar spectral profiles.

## 3 Mathematical formulations of the hybrid similarity measures

Stochastic techniques, such as spectral information divergence, are used to define spectral variation by modelling spectral information as a probability distribution (Chang 2000). In general, stochastic techniques use sample properties and develop spectral criteria such as divergence, probability, etc. to measure dissimilarity between two spectra. Deterministic techniques are based on angle and correlation between two spectra. To combine deterministic and stochastic techniques the algorithms proposed in (Du et al 2004) will be used in this paper.

### 3.1 Spectral correlation angle (SCA)

Given the two spectral signatures $S_i = (s_{il} \ldots s_{il})^T$ and $S_j = (s_{j1}, \ldots, s_{jl})^T$ the Pearsonian correlation coefficient is defined as:

$$r_{s_i, s_j} = \frac{n \sum_1^n s_i s_j - \sum_1^n s_i \sum_1^n s_j}{\sqrt{\left[n \sum_1^n (s_i)^2 - \left(\sum_1^n s_i\right)^2\right]\left[n \sum_1^n s_j^2 - \left(\sum_1^n s_j\right)^2\right]}} \quad (1)$$

where n is the number of spectral bands.

The coefficient is a dimensional index which takes the values anywhere between -1 to 1 and reflects the extent of the linear relationship between the two spectra. To compare with other measures the coefficient converted in to an angle through a formula

$$\mathrm{SCA}(s_i, s_j) = \cos^{-1}\left(\frac{r_{s_i,s_j}+1}{2}\right) \quad \text{in radians ( Bajwa et al 2004)} \tag{2}$$

The SCA takes the values from 0 to 1.570796 radians and is symmetric and invariant to multiplication with positive scalars.

### 3.2 *Spectral angle mapper (SAM)*

SAM is a popular and widely used spectral similarity measure in hyperspectral remote sensing. It calculates spectral similarity by measuring the angle between the spectral signature of two samples, $s_i$ and $s_j$ (Yuhas et al 1992, Kruse et al 1997). The measure determines the similarity between two spectra by calculating the spectral angle between them, treating them as vectors in a space with dimensionality equal to the number of spectral bands used (Kruse et al 1997). The spectral angle has a lower bound of 0 and has values always greater than 1. Unlike the distance metrics it is possible to have a zero spectral angle even when the two vectors are not identical. This technique is relatively insensitive to illumination and albedo effects because the angle between two vectors is invariant with respect to the length of the vectors (Kruse et al 1997). SAM between two spectral signatures with L bands $S_i = (s_{il} \ldots s_{il})^T$ and $S_j = (s_{j1}, \ldots, s_{jl})^T$ is defined as:

$$\mathrm{SAM}(s_i, s_j) = \cos^{-1}\left(\theta_{s_i,s_j}\right) \tag{3}$$

Where $\theta(s_i, s_j) = \left( \dfrac{\sum_{i,j=1}^{L} s_i s_j}{\left[\sum_{i=1}^{L} s_i^2\right]^{\frac{1}{2}} \left[\sum_{j=1}^{L} s_j^2\right]^{\frac{1}{2}}} \right)$

The spectral angle has a maximum value 1.57 and minimum value of 0.

### 3.3 *Spectral information divergence (SID)*

SID is a measure derived from spectral information measure which models the spectral band-band variability as a result of uncertainty caused by randomness. The SID is derived from divergence theory and calculates the probabilistic behaviors between spectral signatures (Van der Meer 2006, Chang 2000). Compared with SAM, which examines the geometrical characters between two spectral signatures or pixel vectors, SID computes the discrepancy between the probability distributions produced by the spectral signatures. Consequently, SID is more effective than SAM in capturing the subtle spectral variability (Chang 2000). SID between two spectral signatures $r_i, r_j$ can be defined as:

$$\mathrm{SID}(r_i, r_j) = D(r_i \| r_j) + D(r_j \| r_i) \tag{4}$$

where

$$D(r_i \| r_j) = \sum_{l=1}^{L} p_l D_l(r_i \| r_j) = \sum_{l=1}^{L} p_l (I_l(r_i) - I_l(r_j)) = \sum_{l=1}^{L} p_l \log_2\left(\frac{p_l}{q_l}\right) \tag{5}$$

And

$$D(r_j \| r_i) = \sum_{l=1}^{L} q_l D_l(r_j \| r_i) = \sum_{l=1}^{L} q_l (I_l(r_j) - I_l(r_i)) = \sum_{l=1}^{L} q_l \log_2\left(\frac{q_l}{p_l}\right) \tag{6}$$

calculated from the probability vectors $\boldsymbol{p}=(p_1,p_2,\ldots,p_L)^T$ and $\boldsymbol{q}=(q_1,q_2,\ldots,q_L)^T$ for the spectral signatures of $s_i$ and $s_j$, where $p_k = \dfrac{s_{ik}}{\sum_{l=1}^{L} s_{il}}$ and $q_k = \dfrac{s_{jk}}{\sum_{l=1}^{L} s_{jl}}$ So the self-information provided by $r_j$ for band $l$ is defined by $I_l(r_i) = -\log_2(p_l)$ and similarly $I_l(r_j) = -\log_2(q_l)$. According to information theory, $D(r_i \| r_j)$ in Equation 4 is called the relative entropy of $r_j$ with respect to $r_i$, which is also known as the Kullback-Leibler information measure (Kullback 1959).

### 3.4 Hybrid measures of spectral information divergence and spectral angle mapper

The SIDSAM mixed measure proposed by (Du et al 2004) to increase discriminability makes two similar spectra even more similar and two dissimilar spectra more distinct. SIDSAM between two spectral signatures $S_i = (s_{il}\ldots s_{il})^T$ and $S_j = (s_{jl},\ldots,s_{jl})^T$ is defined as

$$\text{SIDSAM}_{\tan} = \text{SID}(s_i, s_j) \times \tan(\text{SAM}(s_i, s_j)) \qquad (7)$$

$$\text{SIDSAM}_{\sin} = \text{SID}(s_i, s_j) \times \sin(\text{SAM}(s_i, s_j)) \qquad (8)$$

It should be noted that the cosine is not used in the mixed measure because the cosine calculates the projection of one spectrum along the other one. In this case, taking the cosine of $\text{SAM}(s_i, s_j)$ will reduce discriminability.

### 4 Proposed hybrid similarity measure based on spectral correlation angle and spectral information divergence

In the light of the above for spectral similarity measure, we proposed and developed a new measure SIDSCA which is similar to SIDSAM but with enhanced discriminatory capabilities for separating two similar spectra. The SCA has advantages over SAM (Carvalho et al 2000, Robila 2005) in measuring the spectral properties because of its ability to detect false positive results and is used effectively in classification of hyperspectral images (Carvalho et al 2003 ). The SCA also eliminates negative correlation and maintains the SAM characteristic of minimizing the shading effect resulting in better results. Therefore the new hybrid measure of SIDSCA is expected to improve the discriminatory power as against the existing method of SIDSAM. The new measure SIDSCA between two spectral signatures $S_i = (s_{il}\ldots s_{il})^T$ and $S_j = (s_{jl},\ldots,s_{jl})^T$ is defined as

$$\text{SIDSCA}_{\tan} = SID(s_i, s_j) \times \tan(\text{SCA}(s_i, s_j)) \qquad (9)$$

$$\text{SIDSCA}_{\sin} = \text{SID}(s_i, s_j) \times \sin(\text{SCA}(s_i, s_j)) \qquad (10)$$

The tan and sin in Equation (9), (10) denote the tangent and sine trigonometric functions respectively. The reason for considering tangent and sine trigonometric functions rather than cosine is to calculate the perpendicular distance between the spectra $S_i, S_j$ instead of projection of one spectrum along the other spectra (Du et al 2004).

## 5 Spectral discriminatory measures

Though the spectral similarity measures calculate similarity or dissimilarity between two spectral signatures, but these paired discrimination procedures alone are not enough to discriminate more than two spectral classes. Moreover, as different similarity measures use different units of measurement, it is impossible to evaluate their performance without comparable statistics. Therefore, in order to discriminate a set of spectral classes of different crop species or to determine the relative performance of the measures described above, three statistical algorithms, (i) relative spectral discriminatory probability (RSDPB), (ii) relative spectral discriminatory power (RSDPW) and (iii) relative spectral discriminatory entropy were used.

### 5.1 Relative spectral discriminatory probability (RSDPB)

RSDPB calculates the relative capability of all spectra to be discriminated from others. In general, the higher the probability, the better is the capability of a set of spectra to be discriminated from others. Let $\{s_k\}_{k=1}^{K}$ be $K$ spectral signatures in the set $\Delta$, which can be considered as a database, and t be any specific target spectral signature to be identified using $\Delta$ (Chang 2000). The definition of the RSDPB of all $s_k$ in $\Delta$ relative to t is:

$$P_{t,\Delta}^{m}(k) = \frac{m(t, s_k)}{\sum_{j=1}^{L} m(t, s_j)} \quad \text{for } k=1,\ldots,K \quad (11)$$

Where $\sum_{j=1}^{K} m(t, s_j)$ is the normalization constant and $m(t, s_k)$ is any of the defined spectral similarity measures

We have considered $t$ as pure spectra and also a mixture of two or more crop species whose reflectances are combined in a linear proportion.

The resulting probability vector $P_{t,\Delta}^{m} = (P_{t,\Delta}^{m}(1), P_{t,\Delta}^{m}(2), \ldots, P_{t,\Delta}^{m}(k))^T$ is the RSDPB of $\Delta$ with respect to $t$ or spectral discriminatory probability vector of $\Delta$ relative to $t$. Then, using Equation (11) we can identify $t$ via $\Delta$ by selecting the one with the smallest relative spectral discriminability probability. If there is a tie, either one can be used to identify $t$. Through RSDPB we see the normalized distance measure. Given the target and the reference spectra, we decide that the target matches the spectra with the smallest RSDPB value.

### 5.2 Relative spectral discriminatory entropy (RSDE)

Using a selective set of spectral signatures, $\Delta = \{s_k\}_{k=1}^{K}$, we can further define the relative spectral discriminatory entropy (RSDE) measure of a spectral signature t with respect to the set $\Delta$, as $H_{RSDE}(\mathbf{t}; \Delta)$ given by

$$H_{RSDE}(t; \Delta) = -\sum_{k=1}^{K} P_{t,\Delta}^{m}(k) \log_2\left(P_{t,\Delta}^{m}(k)\right) \quad (12)$$

Equation (12) provides an uncertainty measure of identifying $t$ resulting from using $\Delta = \{s_k\}_{k=1}^{K}$. The measure is seen as a way to analyze the uncertainty with respect to the match between $t$ and reference spectra. A larger entropy value indicates a higher degree of uncertainty with respect to *t*. The lower the entropy value, the higher the chance the targets will be correctly matched (Chang 2000).

### 5.3 *Relative spectral discriminatory power (RSDPW)*

RSDPW lies in calculating how well one spectral vector can be distinguished (discriminated) from another spectral vector, relative to a reference spectral vector (Van der Meer 2006, Chang 2000). Given $m(.,.)$ is a spectral measure, $d$ is the reference spectral signature, and $s_i$ and $s_j$ are the spectral signatures or pair of pixel vectors, the RSDPW of $m(.,.)$ represented by $\Omega(s_i, s_j; d)$ is:

$$\Omega(s_i, s_j; d) = \max\left\{\frac{m(s_i, d)}{m(s_j, d)}, \frac{m(s_j, d)}{m(s_i, d)}\right\} \quad (13)$$

The $\Omega(s_i, s_j; d)$ defined by Equation(13) provides a quantitative index of spectral discrimination capability of a specific hyperspectral measure $m(.,.)$ between two spectral signatures $s_i, s_j$ relative to d. Obviously, the higher the $\Omega(s_i, s_j; d)$ is, the better discriminatory power the $m(.,.)$. In addition, $\Omega(s_i, s_j; d)$ is symmetric and bounded below by one, i.e., $\Omega(s_i, s_j; d) \geq 1$ with equality if and only if $s_i = s_j$

## 6 Results

The data used in the following experiments is obtained using the protocols described in section 2. The reflectance spectra belonging to the five crop species of *Vigna* genus namely, blackgram, greengram, horsegram, cowpea and ricebean are considered for analysis. The correlation coefficient between the crop species is computed and plotted (figure 2) and its relationship with SAM and SCA is carried out.

[Include Figure 2 here]

The SAM and SCA values are found to vary with correlation coefficient. The lowest correlation value of 0.7 between horsegram-cowpea in 400nm-700nm spectral range has produced a value 0.33 by SAM when compared to 0.54 produced by SCA. The highest correlation coefficient 0.998 between greengram-horsegram in 1510nm-2300nm spectral region has resulted in a value of 0.31 by SAM and 0.21 by SCA. In general lower the correlation between the two species higher is the similarity value produced by SCA than SAM in all the spectral regions except 1510nm-2300nm. This is attributed to the sensitivity of SCA to certain selected spectral ranges (Van der Meer 2006).

The magnitude of the similarity values decides the dissimilarity between the crop species (Chang 2000). The higher is the similarity value between the crops species better is the discrimination between them. Therefore the SCA is found to be a better discriminator than SAM based on magnitude of the similarity value. To spectral information divergence is found to enhance the similarity values produced by individual measures therefore the hybrid measures of SIDSCA$_{tan}$ and SIDSAM$_{sin}$ is tabulated in table 1.

[Include Table 1 here]

The magnitude of the similarity value produced by SIDSAM$_{tan}$ and SIDSCA$_{tan}$ are slightly higher than SIDSAM$_{sin}$ and SIDSCA$_{sin}$ respectively which can be seen by rearranging the Equation 3 as SIDSAM$_{tan}$ = $\frac{\text{SIDSAM}_{sin}}{\cos(\text{SAM}(s_i,s_j))}$ = $\frac{\text{SIDSAM}_{sin}}{\theta(s_i,s_j)}$ , and $0 < \theta(s_i,s_j) < 1$ implying that as long as $\theta(s_i,s_j)$ is between 0 and 1 SIDSAM$_{tan}$ has always greater than SIDSAM$_{sin}$. Therefore SIDSAM$_{tan}$ and SIDSCA$_{tan}$ will only be considered for subsequent analysis.

In order to see the performance of the hybrid measures using SID i.e. SIDSAM$_{tan}$ and SIDSCA$_{tan}$ we computed their spectral discriminatory powers. The greengram as reference signature 'd' and horsegram as '$s_i$' and cowpea as '$s_j$' since greengram is more similar to horsegram. The figure 3 shows the spectral discriminatory powers between the crop species with greengram as reference.

[Include Figure 3 here]

The computation of the spectral discriminatory power for 400-700nm spectral range is

$$\Omega^{\text{SIDSAM}_{tan}}(\text{horsegram}, \text{cowpea}; \text{greengram}) = \frac{\text{SIDSAM}_{tan}(\text{cowpea}, \text{greengram})}{\text{SIDSAM}_{tan}(\text{horsegram}, \text{greengram})}$$

$$= \frac{0.04604}{0.00003} \approx 1151 \quad (14)$$

$$\Omega^{\text{SIDSCA}_{tan}}(\text{horsegram}, \text{cowpea}; \text{greengram}) = \frac{\text{SIDSCA}_{tan}(\text{cowpea}, \text{greengram})}{\text{SIDSCA}_{tan}(\text{horsegram}, \text{greengram})}$$

$$= \frac{0.08292}{0.00004} \approx 1716 \quad (15)$$

The equation 14 shows that the discriminatory power of SIDSAM$_{tan}$ hybrid measure to distinguish greengram from horsegram is 1151 times better than to distinguish greengram from cowpea. Compared to SIDSAM$_{tan}$ hybrid measure SIDSCA$_{tan}$ measure is 1.5 times more effective. The correlation between horsegram and greengram in 400nm-700nm region is lower therefore SIDSCA$_{tan}$ spectral discriminatory power is higher than SIDSAM$_{tan}$ when greengram is taken as reference spectra. The details of the computation are given in the table 2.

[Include Table 2 here]

In general the hybrid measure SIDSCA$_{tan}$ is found to have higher relative discriminatory power than SIDSAM$_{tan}$ in 400nm-700nm spectral range.

In order to evaluate which measure is more effective in terms of spectral discrimination, a linear mixture *t* is considered with 0.1 blackgram, 0.2 greengram, 0.5 horsegram, 0.1 cowpea, 0.1 ricebean. The mixture is compared to the spectra in the database $\Delta = \{\text{blackgram}, \text{greengram}, \text{horsegram}, \text{cowpea}, \text{ricebean}\}$ which is made up of five signatures, using spectral discriminatory probability and relative spectral discriminatory entropy measures. The table 3 shows the relative discriminatory probabilities of the crop spectra in the database when compared with the mixture *t*.

[Include Table 3 here]

The ratio of relative spectral discriminatory probability value with $SIDSAM_{tan}$ between $t$ and greengram to $SIDSAM_{tan}$ between $t$ and horsegram is $\frac{0.00053}{0.00382} \approx 0.13989$. In comparison, $SIDSCA_{tan}$ yielded $\frac{0.00050}{0.00318} \approx 0.15843$ which is 1.2 times more effective in identifying $t$ as horsegram than $SIDSAM_{tan}$. The computational results are detailed in the table 4.

[Include Table 4 here]

The relative spectral discriminatory entropy produced by $SIDSCA_{tan}$ is lower when compared to $SIDSAM_{tan}$ in the spectral region 400nm-700nm. Therefore $SIDSCA_{tan}$ has proved to be a better measure in discriminating the *Vigna* species. A decision table given in table 5 based on the RSPDW is developed to find the suitable spectral range for discriminating the crop species of *Vigna*.

[Include Table 5 here]

## 7  Conclusions

A new similarity measure spectral correlation angle is proposed which is based on the spectral correlation mapper (SCM) and is compared with spectral angle mapper in different spectral regions. The lower correlations between the crop spectra have produced higher spectral correlation angles. The spectral correlation angle is found to be higher in all the spectral ranges except 1510nm-2300nm. The higher SCA measurement is attributed to the normalization through averaging process as compared to displacement to zero point in SAM.

A new hybrid technique based on the spectral correlation angle and spectral information divergence is proposed. The new hybrid measure $SIDSCA_{tan}$ has produced higher similarity values than $SIDSAM_{tan}$ in 400nm-700nm spectral range. The proposed hybrid measure $SIDSCA_{tan}$ has also produced a highest discriminatory power between horsegram and cowpea with greengram as reference when compared to $SIDSAM_{tan}$ measure in 400nm-700nm spectral range. Most of the crops have been discriminated using $SIDSCA_{tan}$ in 400nm-700nm spectral range. The RSPDB and RSPDE measures have shown $SIDSCA_{tan}$ as a better discriminator than $SIDSAM_{tan}$ in discriminating horsegram from a mixture of crop spectra belonging to *Vigna* species in 400nm-700nm spectral region.

The $SIDSCA_{tan}$ hybrid measure is found to be a better measure than $SIDSAM_{tan}$ hybrid measure in the 400nm-700nm spectral range. Potential of the proposed $SIDSCA_{tan}$ hybrid measure could be better tested and validated with the spectra of different mixed and pure cover types.

# Acknowledgements

We express our sincere thanks to Dr. V. Jayaraman, Director, National Remote Sensing Centre, for his constant encouragement and guidance. Thanks are due also due to Dr R.S. Dwivedi, Group Director, Land Resources Group, for his suggestions. Thanks are due to officials who have contributed to providing the ground measurements of hyperspectral data.

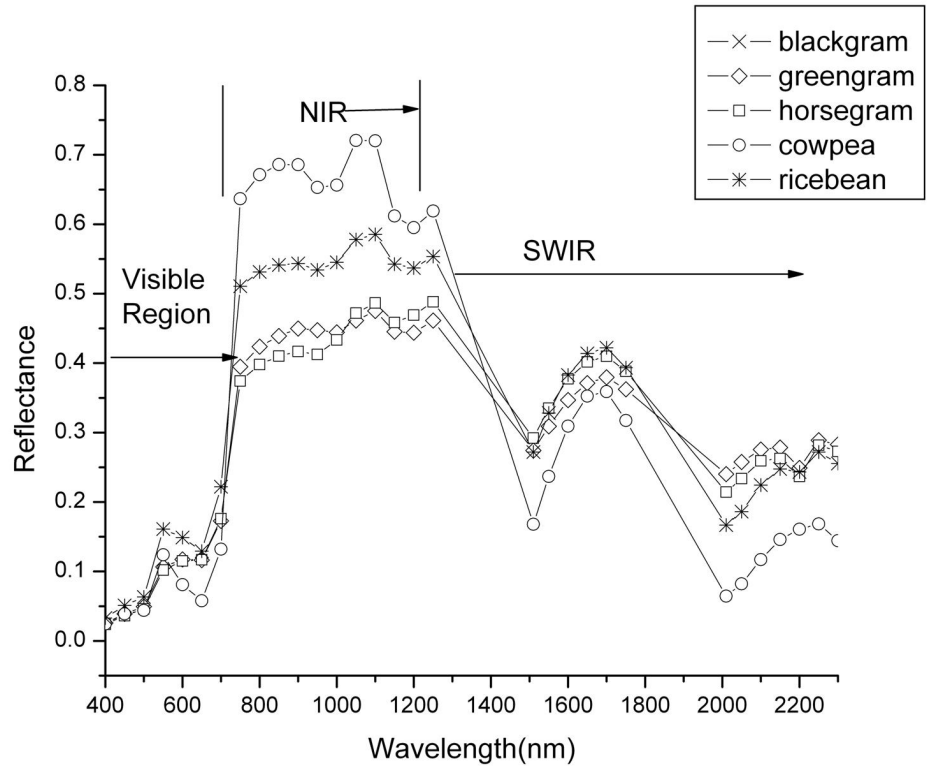

**Figure 1**: *Vigna* crop species laboratory reflectance 400nm-2500nm wavelength

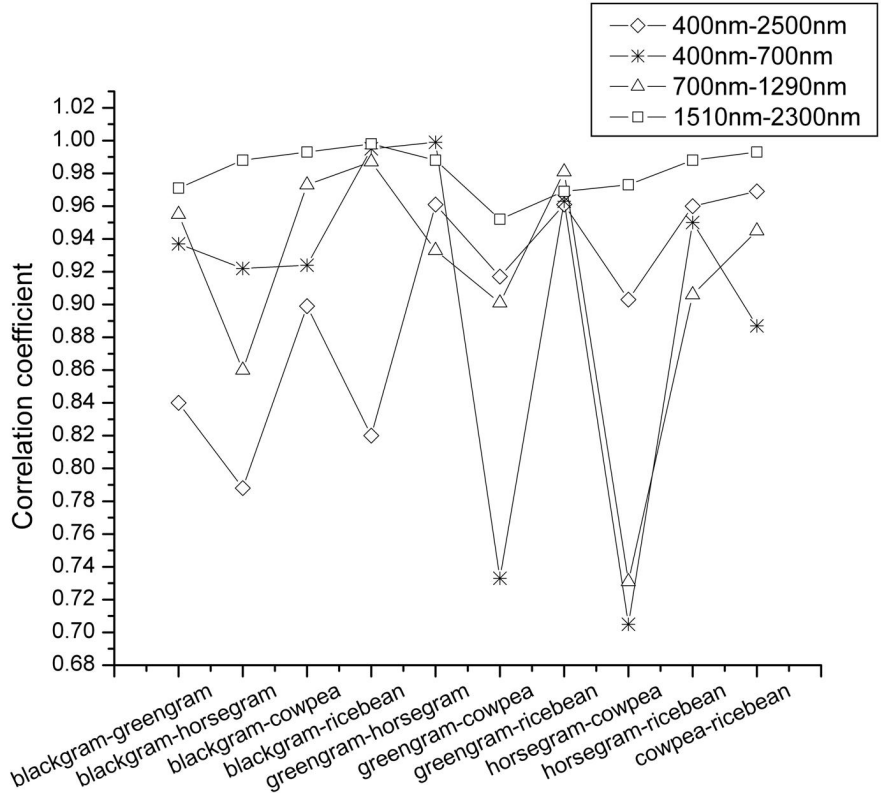

**Figure 2:** Correlation coefficient between the crops in different spectral regions

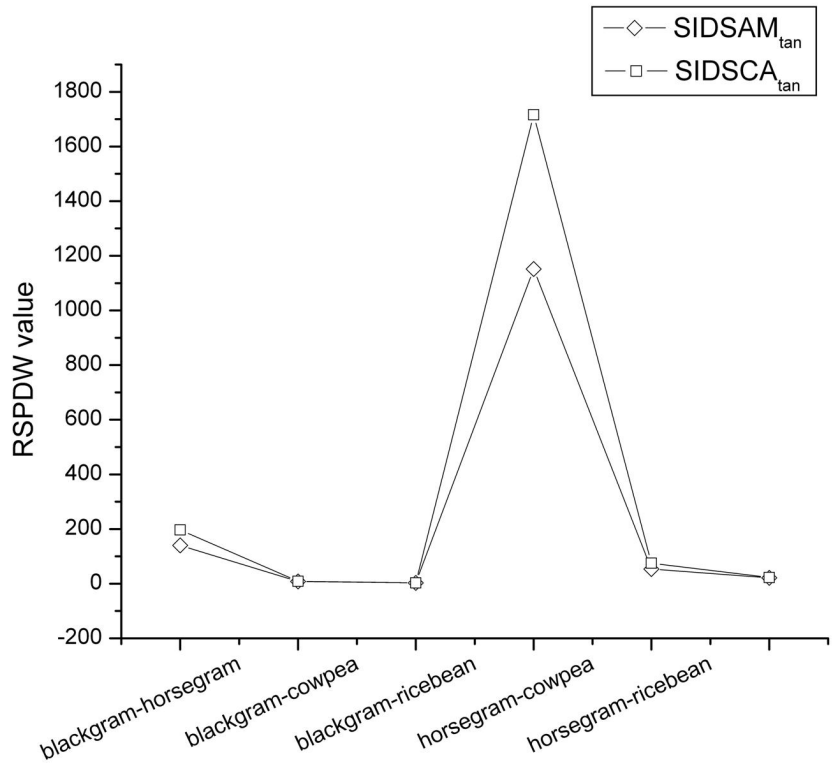

**Figure 3:** Relative spectral discriminatory power between the crop species of *Vigna* in 400nm-700nm range with greengram as reference

Table 1 Similarity Values of crop species in different Spectral range

| Spectral Range<br>Crop Species | 400nm-2500nm | | 400nm-700nm | | 700nm-1290nm | | 1510nm-2300nm | |
|---|---|---|---|---|---|---|---|---|
| | $SIDSAM_{tan}$ | $SIDSCA_{tan}$ | $SIDSAM_{tan}$ | $SIDSCA_{tan}$ | $SIDSAM_{tan}$ | $SIDSCA_{tan}$ | $SIDSAM_{tan}$ | $SIDSCA_{tan}$ |
| blackgram-greengram | $6.3 \times 10^{-2}$ | $8.22 \times 10^{-2}$ | $5.61 \times 10^{-3}$ | $9.51 \times 10^{-3}$ | $2.10 \times 10^{-4}$ | $9.60 \times 10^{-4}$ | $4.03 \times 10^{-2}$ | $2.51 \times 10^{-2}$ |
| blackgram-horsegram | $7.44 \times 10^{-2}$ | $1.01 \times 10^{-1}$ | $9.08 \times 10^{-3}$ | $1.47 \times 10^{-2}$ | $7.60 \times 10^{-4}$ | $3.90 \times 10^{-3}$ | $1.68 \times 10^{-2}$ | $9.02 \times 10^{-3}$ |
| blackgram-cowpea | $2.44 \times 10^{-2}$ | $2.64 \times 10^{-2}$ | $6.20 \times 10^{-3}$ | $1.07 \times 10^{-2}$ | $9.00 \times 10^{-5}$ | $4.00 \times 10^{-4}$ | $8.40 \times 10^{-4}$ | $1.09 \times 10^{-3}$ |
| blackgram-ricebean | $4.00 \times 10^{-2}$ | $5.24 \times 10^{-2}$ | $4.50 \times 10^{-4}$ | $5.40 \times 10^{-4}$ | $2.20 \times 10^{-4}$ | $5.70 \times 10^{-4}$ | $6.14 \times 10^{-3}$ | $2.22 \times 10^{-3}$ |
| greengram-horsegram | $3.99 \times 10^{-3}$ | $5.91 \times 10^{-3}$ | $4.00 \times 10^{-5}$ | $5.00 \times 10^{-5}$ | $1.90 \times 10^{-4}$ | $9.90 \times 10^{-4}$ | $7.50 \times 10^{-4}$ | $1.06 \times 10^{-3}$ |
| greengram-cowpea | $1.02 \times 10^{-1}$ | $8.78 \times 10^{-2}$ | $4.60 \times 10^{-2}$ | $8.29 \times 10^{-2}$ | $7.40 \times 10^{-4}$ | $3.38 \times 10^{-3}$ | $5.96 \times 10^{-2}$ | $4.11 \times 10^{-2}$ |
| greengram-ricebean | $8.29 \times 10^{-3}$ | $9.38 \times 10^{-3}$ | $2.14 \times 10^{-3}$ | $3.63 \times 10^{-3}$ | $3.00 \times 10^{-5}$ | $1.70 \times 10^{-4}$ | $4.67 \times 10^{-3}$ | $5.91 \times 10^{-3}$ |
| horsegram-cowpea | $9.29 \times 10^{-2}$ | $8.88 \times 10^{-2}$ | $5.85 \times 10^{-2}$ | $1.04 \times 10^{-1}$ | $2.34 \times 10^{-3}$ | $1.22 \times 10^{-2}$ | $2.87 \times 10^{-2}$ | $1.92 \times 10^{-2}$ |
| horsegram-ricebean | $6.89 \times 10^{-3}$ | $8.57 \times 10^{-3}$ | $3.81 \times 10^{-3}$ | $6.24 \times 10^{-3}$ | $3.60 \times 10^{-4}$ | $1.83 \times 10^{-3}$ | $6.80 \times 10^{-4}$ | $1.06 \times 10^{-3}$ |
| cowpea-ricebean | $2.77 \times 10^{-2}$ | $2.40 \times 10^{-2}$ | $1.27 \times 10^{-2}$ | $2.20 \times 10^{-2}$ | $6.80 \times 10^{-4}$ | $2.46 \times 10^{-3}$ | $1.15 \times 10^{-2}$ | $5.52 \times 10^{-3}$ |

Table 2 RSPDW comparison of the measures SIDSCA and SIDSAM

| Measure | cowpea – greengram | horsegram-greengram | RSPDW | Ratio of measures |
|---|---|---|---|---|
| $SIDSCA_{tan}$ | $8.29 \times 10^{-2}$ | $4.00 \times 10^{-5}$ | $(8.29 \times 10^{-2}/4.00 \times 10^{-5})=1716$ | $1716/1151=1.5$ |
| $SIDSAM_{tan}$ | $4.60 \times 10^{-2}$ | $3.00 \times 10^{-5}$ | $(4.60 \times 10^{-2}/3.00 \times 10^{-5})=1151$ | |

Table 3 Relative Discriminatory Probabilities and Entropy measure with mixture $t$ as 0.15 blackgram, 0.05 greengram, 0.7 horsegram, 0.06 cowpea and 0.04 ricebean in 400nm-700nm spectral range

| Measure | blackgram | greengram | horsegram | cowpea | ricebean | Entropy |
|---|---|---|---|---|---|---|
| SIDSAM | $8.81 \times 10^{-2}$ | $5.30 \times 10^{-4}$ | $3.82 \times 10^{-3}$ | $8.81 \times 10^{-1}$ | $2.69 \times 10^{-2}$ | $6.47 \times 10^{-1}$ |
| SIDSCA | $8.25 \times 10^{-2}$ | $5.00 \times 10^{-4}$ | $3.18 \times 10^{-3}$ | $8.88 \times 10^{-1}$ | $2.57 \times 10^{-2}$ | $6.17 \times 10^{-1}$ |

Table 4 Computation for finding the best measures for discrimination of t spectral library

| Measure | $t$(greengram) | $t$(horsegram) | $t$(greengram)/$t$(horsegram) | Ratio of measures |
|---|---|---|---|---|
| SIDSCA | $5.00 \times 10^{-4}$ | $3.18 \times 10^{-3}$ | $1.57 \times 10^{-1}$ | $(1.57 \times 10^{-1})/(1.39 \times 10^{-1})=1.13$ |
| SIDSAM | $5.30 \times 10^{-4}$ | $3.82 \times 10^{-3}$ | $1.39 \times 10^{-1}$ | |

Table 5 Decision table for selecting the reference spectra, spectral range and measure which has produced highest discriminatory power

| | greengram | | | horsegram | | |
|---|---|---|---|---|---|---|
| | Reference | Range | measure | Reference | Range | measure |
| blackgram | horsegram | 400nm-700nm | SIDSCA | greengram | 400-700nm | SIDSCA |
| greengram | | | | ricebean | 700nm-1290nm | SIDSAM |
| | cowpea | | | ricebean | | |
| | Reference | Range | measure | Reference | Range | measure |
| blackgram | ricebean | 400nm-700nm | SIDSCA | horsegram | 1510nm-2300nm | SIDSAM |
| greengram | horsegram | 400nm-700nm | SIDSCA | horsegram | 400nm-700nm | SIDSCA |
| horsegram | greengram | 400nm-700nm | SIDSCA | greengram | 400nm-700nm | SIDSCA |
| cowpea | | | | horsegram | 1510nm-2300nm | SIDSAM |